\documentstyle[12pt]{article}

\begin{document}
\setlength{\baselineskip}{0.9cm}
\def\beq{\begin{equation}}
\def\eeq{\end{equation}}
\def\beqa{\begin{eqnarray}}
\def\eeqa{\end{eqnarray}}
\def\S{{\rm S}}

\begin{titlepage}
\title{\bf BRST Cohomology and Renormalizability of Quantum Gravity near
Two Dimensions} 
\author{{\sc Yoshihisa Kitazawa}, {\sc Rie Kuriki and Katsumi Shigura}} 
\maketitle
\thispagestyle{empty}
\begin{center}

\vfill

{\it Department of Physics, Tokyo Institute of Technology\\ Oh-Okayama,
Meguro-ku,\\
Tokyo 152, Japan}

\vfill

\small{kitazawa@th.phys.titech.ac.jp,~r-kuriki@th.phys.titech.ac.jp,
shigura@th.phys.titech.ac.jp}\\

\vfill

{\bf Abstract}
\end{center}
\begin{quote}
We discuss
the renormalizability of quantum gravity near two dimensions based on the
results
obtained by a computation of the BRST-antibracket cohomology in the space
of local 
functionals of the fields and antifields. We justify the assumption
on the general structure of the counterterms which have been used in the
original proof of
renormalizability of the quantum gravity near two dimensions. 
\end{quote}

\vfill
\vbox{
\hfill TIT/HEP-364 \null\par
}\null

\end{titlepage}

\section{Introduction}

Recently,
the renormalizability of the quantum gravity near two dimensions has been
actively studied
\cite{Kitazawa1}\cite{Kitazawa2}\cite{Kitazawa3}. The authors in
ref.\cite{Kitazawa2} have proven that
all necessary counterterms
which may be required
in the perturbative calculation can be supplied by the bare action which
is invariant under the full diffeomorphisms. 
The gauge invariance imposes very strong constraints on the possible
counterterms which are required to cancel the short distance divergences.
However the conformal invariance is well known to be anomalous
in two dimensional quantum gravity. 
Nevertheless the structure of the conformal anomaly
also imposes strong constraints on the possible divergences.
In fact it is assumed that the possible counterterm which is compatible
with the anomaly is unique in  \cite{Kitazawa2} .
proof of
To provide the justification for this crucial assumption on the uniqueness
of the counterterms 
is a main motivation of this paper. 

The authors in the ref.\cite{Kitazawa2}
start with the tree level action which
possesses the volume preserving
diffeomorphism invariance.
In order to recover the full diffeomorphisms invariance,
they further require that
the theory is independent of the background metric. As they argue,
this requirement leads us to search
a theory which is conformally invariant
with respect to the background metric.
Obviously the Einstein action is such a theory. They finally conjecture that
the requirement of the background
independence leads us uniquely to the
Einstein action
which is invariant under the diffeomorphisms, as the bare action. We will
try to justify
the conjecture on the ground of mathematical and model independent analysis
in this paper.

It has been recognized
that the BRST method provides us one of the most powerful tools for
quantizing theories
with local gauge symmetries \cite{BRS}.
The original BRST method has been extended further. One of the extended
BRST quantization methods is the field-antifield formalism developed by Batalin-Vilkovisky \cite{BV1} \cite{BV2}. 
In this formulation, an odd symplectic structure which is called
antibracket plays an important role.
The BRST-antibracket
cohomology
of the two-dimensional Weyl invariant
gravity theory has been completely
computed
in the space of local functionals of the fields and antifields in
ref.{\large }\cite{Brandt1}.
The starting point of the analysis used in that work is the field content,
and the symmetry
transformations.
These include the diffeomorphisms and the Weyl transformations. 

These symmetries are realized on the scalar matter fields and on the
two-dimensional metric:
\beqa
&& \delta X^\mu = \xi^\rho \partial_\rho X^\mu, \nonumber\\ 
&& \delta g_{\alpha\beta} = \xi^\rho \partial_\rho g_{\alpha\beta} +
g_{\alpha\rho}\partial_\beta \xi^\rho
+ g_{\beta\rho}\partial_\alpha \xi^\rho
+ Cg_{\alpha\beta}, \nonumber\\
&& \delta \xi^\alpha = \xi^\rho \partial_\rho \xi^\alpha, \nonumber\\ 
&& \delta C = \xi^\rho \partial_\rho C,
\eeqa 
where $C$ and $\xi^\alpha$
are respectively
the Weyl and diffeomorphism ghosts and $\delta$ is the BRST transformation.

%
provided by
diffeomorphism ghosts, 
%
In the language of the BRST cohomology,
generally, the anomalies (Weyl anomaly, gauge anomaly, etc...) are
represented by cohomology classes with
ghost number one. It can be used to compute other classes with the
different values of the ghost number.
In particular the class with ghost number zero is interesting since it
contains generic classical actions.
This gives the possibility
to construct the most general classical action by computing the class with
ghost number zero for the 
desired field content and gauge invariances. 
%
polynomial in derivatives of all
%

Here we briefly summarize
the results \cite{Brandt1} which will be mainly used in this paper.
In the ghost number zero sector,
the BRST cohomology without antifields
determines the most general classical action: 
\beq
\frac{1}{2} \int d^2 x (\sqrt{g} g^{\alpha\beta} \partial_\alpha X^\mu
\partial_\beta X^\nu G_{\mu\nu}(X)
+ \epsilon^{\alpha\beta} \partial_\alpha X^\mu \partial_\beta X^\nu
B_{\mu\nu}(X)),
\label{ghost0}
\eeq
where $G_{\mu\nu}(X), B_{\mu\nu}(X)$ are arbitrary functions of $X^\mu$
that are respectively symmetric and
antisymmetric 
under $\mu \leftrightarrow \nu$, and $\epsilon^{\alpha\beta}$ is
the constant antisymmetric tensor.
The BRST cohomology in the ghost number
one sector gives the candidates of anomalies. It seems reasonable to assume
that the anomalies are left-right symmetric since the action (\ref{ghost0})
is left-right symmetric. 
Accordingly the candidates of anomalies
in two dimensional conformal gravity are
\beqa
&& \int d^2 x C\sqrt{g} R,
\label{ghost1}\\
&& \frac{1}{2} \int d^2 x C(\sqrt{g} g^{\alpha\beta} \partial_\alpha X^\mu
\partial_\beta X^\nu f_{(\mu\nu)}(X)
+ \epsilon^{\alpha\beta} \partial_\alpha X^\mu \partial_\beta X^\nu
f_{[\mu\nu]}(X)),
\label{ghost2}
\eeqa
where
$f_{(\mu\nu)}(X)$ and $f_{[\mu\nu]}(X)$ are arbitrary functions of the
matter fields $X^\mu$. 
symmetrization of $\mu,\nu$.
We recall here that the diffeomorphism
anomaly and the Weyl anomaly are
cohomologically equivalent.
Therefore we have a choice to
respect either the diffeomorphism invariance or the conformal invariance but
not the both.
We choose to respect 
the diffeomorphism invariance when we quantize the theory.

\medskip

The organization of this paper is as follows. In section two, we set up a
model which can preserve
the diffeomorphism
and the Weyl invariance in the $2+\varepsilon$ dimensions. Then we briefly
review
the proof of the renormalizability
of $2+\varepsilon$ dimensional quantum
gravity. In section three,
we prove the uniqueness of
the counterterms based on the BRST cohomology which justifies the
assumption used in \cite{Kitazawa2}.
This proof of the uniqueness of the counterterms is the essential part of
this paper.
We conclude in section four with discussions. 

\section{Quantum gravity in $2+\varepsilon$ dimensions} 

The classical action
which possesses the diffeomorphism
and the Weyl invariance in the $2 + \varepsilon$ dimensions is described by
\beqa
S &=& \frac{1}{2}
\int d^{2 + \varepsilon} x \sqrt{g} (\partial_\alpha X^\mu \partial_\beta
X^\nu g^{\alpha\beta} G_{\mu\nu}(X) \nonumber \\ 
&& ~~~~~~~~~~ {}
+ \epsilon^{ab} e^\alpha_a e^\beta_b \partial_\alpha X^\mu \partial_\beta
X^\nu B_{\mu\nu}(X))
( 1 + \frac{1}{2}\sqrt{\frac{\varepsilon}{2(D - 1)}} \psi )^2 \nonumber \\
&+& \!\! \int d^{2 + \varepsilon} x \sqrt{g} \{ R ( 1 +
\sqrt{\frac{\varepsilon}{2(D - 1)}}\psi 
+ \frac{\varepsilon}{8(D - 1)}\psi^2 ) - \frac{1}{2} \partial_\mu \psi
\partial_\nu \psi g^{\mu\nu} \}, \nonumber \\ 
&&
\label{2eaction}
\eeqa
where $X^\mu$
$(\mu=0,\ldots,D-1)$
is a set of scalar matter fields, and $\psi$ is another scalar field.
$G_{\mu\nu}$ and $B_{\mu\nu}$ $(\mu=0,\ldots,D-1)$ are, respectively,
arbitrary functions of the $X^\nu$ satisfying 
\beqa
G_{\mu\nu}=G_{\nu\mu},~~B_{\mu\nu}=-B_{\nu\mu}, 
\eeqa 
and $g^{\alpha\beta}$ are
the metric on the $2+\varepsilon$ dimensional world sheet.
The "twei"-bein field which
is a connection between 
the world sheet and the tangent space is denoted by
$e^\alpha_a$, the index $a$ refers
to the local Lorentz transformation on
the tangent space.
$\epsilon^{ab}$ is a constant
antisymmetric tensor on the $2+\varepsilon$ dimensional tangent space of
the world sheet
\cite{Tseytlin}. 

At the limit of
$\varepsilon \to 0$, the form of
(\ref{2eaction}) agrees with
the form of (\ref{ghost0}) which is
obtained by considering the BRST cohomology
with ghost number zero.
In this identification, the following Einstein term 
\beq
\int d^2 x \sqrt{g} R
\eeq
can be ignored as the surface term.

\medskip

Next we recall
the standard perturbative evaluation of the effective action by $\hbar$
expansion.
Due to the BRST invariance,
the effective action $\Gamma$ 
satisfies the following Ward-Takahashi identity:
\beq
\Gamma * \Gamma = 0 ,
\label{WT} 
\eeq
It is well-known that
the Eq.(\ref{WT}) imposes strong restriction on the possible divergences
which may appear
in the perturbative evaluation of the effective action. 

The effective action may be expanded in $\hbar$ as %
\beq
\Gamma = \S + \hbar
\Gamma^{(1)} + \hbar^2\Gamma^{(2)} + \cdots , \label{gamma} 
\eeq %
where $\S$ is the classical action
and $\Gamma^{(n)}$ $(n=1,2,\ldots)$ are
the quantum corrections of the theory.
Substituting Eq.(\ref{gamma}) into Eq.(\ref{WT}), one obtains 
\beqa
&& \S * \S = 0, \label{order0}\\
&& \S * \Gamma^{(1)} = 0,
\label{level1} \\
&& \Gamma^{(1)} * \Gamma^{(1)} + 2\, \S * \Gamma^{(2)} = 0, \\ 
&& \cdots\cdots. \nonumber
\eeqa
Here the equation (\ref{order0}), $\S * \S = 0$,
is trivially satisfied
if the classical action $S$ preserve the gauge symmetries.
The equation (\ref{level1}) constrains the possible structure of the
counter terms at the one loop level.
In fact we prove that the structure of the counter terms are 
unique in the next section. As it will be shown there, we can iterate this
procedure
order by order in the $\hbar$ expansion to all orders
and the proof of the renormalizability follows inductively.

\section{Uniqueness of the counterterms} 

Our task is to determine the possible counterterms of the theory
which is consistent with the equation (\ref{level1}).
The effective action $\Gamma^{(1)}$
at the one-loop may be divided into the
finite and infinite parts,
which are denoted by $\Gamma_{fin.}$
and $\Gamma_{div.}$ respectively:
\beq
\Gamma^{(1)} = \Gamma_{fin.} + \Gamma_{div.} 
\eeq
The infinite part should be understood
as the term which possesses the $1/\varepsilon$ pole in this paper.
The possible structure of $\Gamma_{div.}$ is constrained by the WT identity
due to the following
considerations.

We find that $\S * \Gamma_{div.}$
is finite since $\S * \Gamma_{fin.}$ is
finite.
By using Eq.(\ref{order0}) and the Jacobi identity on the operation $*$, 
we also find that $\S * ( \S * \Gamma_{div.}) = 0$. 
Hence we conclude that
$\S * \Gamma_{div.}$ is nothing else but an anomaly if it is nonvanishing. 
If the gauge symmetry is independent of the regularization parameter,
we can conclude that $\S * \Gamma_{div.}=0$. It is indeed the case for 
QED or QCD. These theories certainly do not possess anomalies.
However our gauge transformation (conformal transformation) depends on the 
$\varepsilon$ explicitly. Therefore we cannot conclude that $\S *
\Gamma_{div.}=0$.
However we can still conclude that $\S * \Gamma_{fin.}$ is
finite and must be consistent with the anomaly.

As we have already known
all anomalies from the BRST cohomological analysis, we can uniquely
determine the possible divergent 
part $\Gamma_{div.}$ as follows:
\beqa
&& \frac{2}{\varepsilon} \int d^{2 + \varepsilon} x \sqrt{g} R,
\label{2eanomaly}\\
&& \frac{1}{\varepsilon} 
\int d^{2 + \varepsilon} x \sqrt{g} (g^{\alpha\beta}
\partial_\alpha X^\mu \partial_\beta X^\nu f_{(\mu\nu)}(X) + \epsilon^{ab}
e^\alpha_a e^\beta_b \partial_\alpha X^\mu \partial_\beta X^\nu
f_{[\mu\nu]}(X)).
\label{matterdependent}
\eeqa
It is because
\beqa
&& \delta\!\left(\frac{2}{\varepsilon}
\int d^{2 + \varepsilon} x \sqrt{g} R \right) = \int d^2 x C\sqrt{g} R, \\
&& \delta\!\left(\frac{1}{\varepsilon}
\int d^{2 + \varepsilon} x \sqrt{g}
(g^{\alpha\beta} \partial_\alpha
X^\mu \partial_\beta 
X^\nu f_{(\mu\nu)}(X) + \epsilon^{ab} e^\alpha_a e^\beta_b
\partial_\alpha X^\mu 
\partial_\beta X^\nu f_{[\mu\nu]}(X))\right) \nonumber \\
&& = \frac{1}{2}
\int d^2 x C(\sqrt{g} g^{\alpha\beta} \partial_\alpha X^\mu \partial_\beta
X^\nu f_{(\mu\nu)}(X)
+ \epsilon^{\alpha\beta} \partial_\alpha X^\mu \partial_\beta X^\nu
f_{[\mu\nu]}(X)),
\eeqa
where $\delta$ means
the BRST transformation.
On the right-hand side of these equations, we have taken the $\varepsilon
\rightarrow 0$ 
limit since such a limit is well defined. Indeed they become the anomalies
which are
shown in the equations (\ref{ghost1}) and (\ref{ghost2}) in such a
limit.

What we have shown here is that these divergences are consistent with the
anomalies.
In order to prove the uniqueness of the counter terms, let us suppose that
there are two different solutions for $\Gamma_{div.}$.
Then the difference of the two solutions $\Delta (\Gamma_{div.})$ 
satisfies $\S * \Delta (\Gamma_{div.})=0$.
Therefore the possible freedom of the solution is the addition of such 
divergences which satisfy  $\S * \Delta(\Gamma_{div.}) = 0 $.
However this freedom is also constrained by the BRST cohomology
analysis with the ghost number zero.
The only such a solution is of the original action type 
as we have listed in the equation (\ref{ghost0}) in the introduction.
These divergences can be canceled
by the renormalization of the couplings
and the fields.
Of course we may also have BRST
trivial divergences.
However they can be 
renormalized by the wave function and the gauge fixing part.
Due to the triviality of the renormalization of the divergences which
satisfy $\S * \Gamma_{div.} = 0 $, 
it is sufficient to consider  
the anomaly type divergences to prove the renormalizability. 
This completes the proof of the uniqueness of the nontrivial counter terms
of the quantum gravity in
$2+\varepsilon$ dimensions.

\medskip

In the remaining part of this paper, we impose the following global
symmetry on the matter $X^\mu$ field:
\beq
X^\mu~~\longrightarrow~~X^\mu~+~C^\mu,
\eeq
where $C^\mu$ is a constant vector.
By imposing the above global symmetry, one can kill the existence of the
counterterms which contain 
arbitrary functionals of 
$X^\mu$, $G_{\mu\nu}$ and $B_{\mu\nu}$. 
In this way,
we restrict 
our considerations to the matter $X^\mu$ independent Weyl anomaly. 

%
%
%
%
%
%

Let us consider the classical action
\beq
\S = \int d^{2 + \varepsilon}\! x \sqrt{g} 
(R \psi^2 \frac{\varepsilon}{8(D - 1)} - \frac{1}{2} \partial_\mu \psi
\partial_ \nu \psi g^{\mu\nu}). 
\label{action}
\eeq
(\ref{action}) is invariant under the following BRST transformations (the
infinitesimal gauge transformations): 
\beqa
&& \delta{g}_{\alpha\beta} = \xi^\rho \partial_\rho g_{\alpha\beta} 
+ g_{\alpha\rho}\partial_\beta \xi^\rho
+ g_{\beta\rho}\partial_\alpha \xi^\rho
+ Cg_{\alpha\beta}, \nonumber \\
&& \delta\xi^\alpha = \xi^\rho \partial_\rho \xi^\alpha,\nonumber\\ 
&& \delta{C} = \xi^\rho \partial_\rho C,\nonumber\\ 
&& \delta\psi = \xi^\rho
\partial_\rho \psi - C \frac{\varepsilon}{4}\psi.
\label{BRST2e}
\eeqa
(\ref{action}) can be transformed into the matter $X^\mu$ independent part
in (\ref{2eaction}) by the following translation: 
\beq
\psi \to \psi + 2\sqrt{\frac{2(D - 1)}{\varepsilon}}. 
\eeq
There emerges 
the following BRST nontrivial divergence at the one loop level: 
\beq
-\frac{2}{\varepsilon}\int d^{2 + \varepsilon}\! x \lambda\sqrt{g} R, 
\label{Einanomaly} 
\eeq
where $\lambda$ is a known constant.
The reason why 
there is such a divergence at the one loop level is that it is
the only possible nontrivial candidate as we have listed in the equation
(\ref{2eanomaly}).
On the other hand, let $\psi \to \psi + (4 / \varepsilon)\tau$ in
(\ref{action}), then
\beq
\int d^{2 + \varepsilon}\! x \sqrt{g}
(R \psi^2 \frac{\varepsilon}{8(D - 1)} - \frac{1}{2} \partial_\mu \psi
\partial_ \nu \psi g^{\mu\nu}) 
+ \tau\int d^{2 + \varepsilon}\! x\psi\sqrt{g} R +
\frac{2}{\varepsilon}\tau^2\int d^{2 + \varepsilon}\! x\sqrt{g} R. 
\label{shiftaction} 
\eeq
If $\tau$ is put to $\sqrt{\lambda}$, the divergence (\ref{Einanomaly}) is
canceled by the last term of (\ref{shiftaction}). 
After the cancellation, we obtain the following action: 
\beq
\S + \tau\S' = \int d^{2 + \varepsilon}\! x \sqrt{g} 
(R \psi^2 \frac{\varepsilon}{8(D - 1)} - \frac{1}{2} \partial_\mu \psi
\partial_ \nu \psi g^{\mu\nu}) 
+ \tau\int d^{2 + \varepsilon}\! x\psi\sqrt{g} R, 
\eeq
This action is the familiar linear dilaton type.
The BRST transformation with respect to $\psi$ field becomes:
\beq
\delta\psi = \xi^\rho\partial_\rho\psi - C\frac{\varepsilon}{4} 
\psi - \tau C.
\label{qBRST}
\eeq
Since $\lambda$ is $O(\hbar )$, we regard $\tau$ to be $O(\sqrt{\hbar} )$.
Therefore one may expand the effective action as follows: 
\beq
\Gamma = \S + \tau\S' + \hbar\Gamma^{(1)} + \hbar\tau\Gamma^{(1)'} 
+ \hbar^2\Gamma^{(2)} + \hbar^2\tau\Gamma^{(2)'} + \cdots. 
\eeq

As the classical action $\S$ and the BRST transformation have the symmetry
$\psi \to -\psi$ and $\tau \to -\tau$, 
the effective action $\Gamma$ is also invariant under them. $\Gamma^{(n)'}$
do not contain the divergences
of the form $(\ref{Einanomaly})$
since they have to contain odd numbers of $\psi$ field due to the discrete
symmetry. 

Let us suppose that $\Gamma$ is renormalizable up to order $\hbar^{n-1}$.
At next order $\hbar^n$, 
$\S * \Gamma^{(n)}$ must be a finite quantity since it is related to the
finite quantity
by the Ward-Takahashi identity . 
On the other hand,
$\Gamma^{(n)}$ may be generally
split into a finite part $\Gamma^{(n)}_{fin.}$ and an infinite part
$\Gamma^{(n)}_{div.}$. 
Notice that $\S * \Gamma^{(n)}_{div.}$ becomes a finite quantity since $\S
* \Gamma^{(n)}$ is finite.
Here one concludes again that
$\S * \Gamma^{(n)}_{div.}$ should be identified with an anomaly if it is
nonvanishing
since 
\beq
\S * (\S * \Gamma^{(n)}_{div.}) = 0 .\nonumber 
\eeq
$\Gamma^{(n)}_{div.}$ should correspond to (\ref{div}). 
Therefore we 
conclude that the nontrivial solution for $\Gamma^{(n)}_{div.}$ is 
also of $(\ref{Einanomaly})$ type.

Let us transform the action and the BRST transformations of the fields as
follows,
\beqa
&& \int d^{2 + \varepsilon}\! x \sqrt{g} 
(R \psi^2 \frac{\varepsilon}{8(D - 1)} - \frac{1}{2} \partial_\mu \psi
\partial_ \nu \psi g^{\mu\nu}) \nonumber\\ 
&& + \ (\tau + \tau_2 + \cdots + \tau_n)\int 
d^{2 + \varepsilon}\! x\psi\sqrt{g} R \nonumber\\ 
&& + \ \frac{2}{\varepsilon} (\tau + \tau_2 + \cdots + \tau_n)^2 \int d^{2
+ \varepsilon}\! x \sqrt{g} R,
\eeqa
\beq
\delta{\psi} = \xi^\rho\partial_\rho\psi - C\frac{\varepsilon}{4} 
\psi - (\tau + \tau_2 + \cdots + \tau_n)C 
\eeq
where $\tau_n \sim \hbar^{n - 1}\tau$.

Supposing we find the BRST nontrivial divergence of the following form:
\beq
\hbar^n\Gamma^{(n)}_{div.} = -\frac{4}{\varepsilon}\lambda\int 
d^{2 + \varepsilon}\! x\sqrt{g} R,
\eeq
it  can be renormalized by putting
\beq
\tau_n = \frac{\lambda - \tau_2\tau_{n-1} \cdots}{\tau}, 
\eeq
at order $\hbar^n$.
>From these considerations,
the renormalizability of the $2+\varepsilon$ gravity theory has been proved
recursively.

\section{Conclusions and discussions}
We have given the justification of the assumption which played an important
role in the original 
proof of the renormalizability of $2+\varepsilon$ dimensional quantum
gravity \cite{Kitazawa1} \cite{Kitazawa2}.
The analysis of the BRST antibracket cohomology which is independent of the
model under given gauge 
symmetries and field content allows us to guarantee the uniqueness of the
form of the counterterms.

In this paper,
we have applied the candidates of anomalies in the two-dimensional Weyl
invariant gravity theory 
\cite{Brandt1} to the construction of the counterterms for the $2 +
\varepsilon$ dimensional gravity theory. 
By imposing the translation symmetry of $X^\mu$ and considering the matter
$X^\mu$ independent part, we have proved
the uniqueness of the counterterms which have been necessary to the proof of
the renormalizability of the $2+\varepsilon$ dimensional gravity theory.
Within this framework, we can show that
the only possible BRST nontrivial divergence is
(\ref{Einanomaly})
by using the analysis of the BRST antibracket cohomology.

Our investigation is still confined to the vanishing cosmological
constant case. We may adopt the standard strategy to renormalize the
cosmological constant operator
for infinitesimally small cosmological constant. 
The cosmological constant operator explicitly breaks the conformal
invariance in two
dimensions and hence is not usually considered in the BRST analysis of Weyl
invariant gravity. 
Nevertheless 
we can construct such an operator which is invariant under the
BRST 
transformation in 
$2+\varepsilon$ dimensions (\ref{BRST2e}) and (\ref{qBRST}):
\beq
\int d^{2+\varepsilon} x \sqrt{g}(1+{\varepsilon\over 4\tau}\psi )^{2D\over
\varepsilon}
\sim \int d^{2+\varepsilon} x \sqrt{g} exp(\psi / \tau)
\eeq
Since the quantum correction $\tau$ appears in the denominator in this
expression,
we need to sum the quantum corrections to all orders to obtain the scaling
exponents of the cosmological constant operator and general gravitationally
dressed
operators. In fact it has been successful to reproduce the exact scaling
exponents of two dimensional quantum gravity in this way.
It has been argued that the bare cosmological constant operator can be
renormlized 
in this form to all orders in \cite{Kitazawa2}. 
Since this expression is unique to respect the BRST invariance of
(\ref{BRST2e}) and (\ref{qBRST}),
we believe 
that such an argument can also be justified rigorously in the near 
future.

anomalies 
renormalizability.

\newpage



\begin{thebibliography}{99}
\bibitem{Kitazawa1}
Y. Kitazawa, Nucl. Phys. {\bf B453} (1995) 477. \bibitem{Kitazawa2} 
H. Kawai, Y. Kitazawa and M. Ninomiya, Nucl. Phys. {\bf B467} (1996) 313. 
\bibitem{Kitazawa3}
T. Aida, Y. Kitazawa, preprint hep-th/9609077, to appear in Nucl. Phys. B. 
\bibitem{BRS} C. Becchi, A. Rouet~and~R. Stora,
\newline
Phys. Lett.{\bf B52} (1974) 344;
Comm. Math. Phys.{\bf 42} (1975) 127;
Ann. Phys. {\bf 98} (1976) 287.
\bibitem{BV1}
I. A. Batalin and G. A. Vilkovisky, Phys. Lett. {\bf B102} (1981) 27. 
\bibitem{BV2}
M. Henneaux, C. Teitelboim,
{\it Quantization of Gauge Systems},
Princeton University Press, Princeton 1992. 
\newline
J. Gomis, J. Paris~and~S. Samuel, Phys. Rep. {\bf 259} (1995) 1. 
\newline W. Troost~and~A. Van~Proeyen,
{\it An Introduction to Batalin-Vilkovisky Lagrangian Quantization}, Leuven
Notes in Math. Theor. Phys., in preparation. 
\bibitem{generalBV1} F. Brandt, preprint hep-th/9604025.
\bibitem{generalBV2}
G. Barnich, F.Brandt and M.Henneaux, Commun. Math. Phys. {\bf 174} (1995) 57. 
\bibitem{generalBV3}
G. Barnich, F. Brandt 
and M. Henneaux, Commun. Math. Phys. {\bf 174} (1995) 93. 
\bibitem{Brandt1}
F. Brandt, W. Troost and A. V. Proeyen, Nucl. Phys. {\bf B464} (1996) 353. 
\bibitem{Brandt2}
F. Brandt, W. Troost and A. V. Proeyen,
Nucl. Phys. {\bf B455} (1995) 357.
\bibitem{Tseytlin}
A.A. Tseytlin, Nucl.Phys. {\bf B294} (1987) 383.
\end{thebibliography}
\end{document}